\documentclass[
 ,final  
 ,numberedheadings 
 ]
 {aipproc}

\usepackage{theorem,amsmath,amssymb,mathrsfs}
\usepackage{float,supertabular}
\usepackage{eucal}
\usepackage[hypertex,linkcolor=red]{hyperref}
\usepackage[normalem]{ulem}
\layoutstyle{6x9}

\def\index#1{}
\def\<{\langle}\def\>{\rangle}
\def\Tr{\operatorname{Tr}}\def\:{\hbox{\bf :}}
\renewcommand{\geq}{\geqslant}\renewcommand{\leq}{\leqslant}

\def\Reals{\mathbb R}\def\Cmplx{\mathbb C}
\def\map#1{{\mathscr{#1}}}

\def\set#1{{\sf #1}}\def\alg#1{{\mathcal #1}}\def\aA{\alg{A}}
\def\aI{\alg{I}}

\def\transpPhi#1{\tau_\Phi(#1)}\def\transpOm#1{\tau_\Omega(#1)}
\def\transp#1{{#1}^{\, t}}

\def\adPhi{\operatorname{ad_\Phi}}
\def\adOm{\operatorname{ad_\Omega}}

\def\dim{\operatorname{dim}}

\def\Span{\set{Span}}

\def\Bnd#1{\set{B(#1)}}
\def\sH{\set{H}}

\def\qed{$\,\blacksquare$\par}
\def\eg{e. g. }\def\ie{i. e. }
\def\n#1{|\!|#1|\!|}
\def\dag{\dagger}\def\eff{{\rm eff}}
 
\newtheorem{postulate}{Postulate}

\theoremstyle{remark}
\theoremstyle{example}

\def\trnsfrm#1{\mathscr #1}
\def\tA{\trnsfrm A}\def\tB{\trnsfrm B}\def\tC{\trnsfrm C}
\def\tS{\trnsfrm S}
\def\tI{\trnsfrm I}\def\tT{\trnsfrm T}\def\tU{\trnsfrm U}\def\tM{\trnsfrm M}\def\tX{\trnsfrm X}

\def\tZ{\trnsfrm Z}\def\tF{\trnsfrm F}

\def\AA{\mathbb A}\def\AB{\mathbb B}\def\AL{\mathbb L}
\def\cA{{\underline{\tA}}}\def\cB{\underline{\tB}}\def\cC{\underline{\tC}}\def\cT{\underline{\tT}}
\def\cI{{\underline{\tI}}}
\def\cX{{\underline{\tX}}}

\def\kk{\rangle\!\!\rangle}\def\bb{\langle\!\!\langle}
\def\Stset{{\mathfrak S}}
\def\Trnset{{\mathfrak T}}
\def\Cntset{{\mathfrak P}}
\begin{document}
\title{Operational axioms for C${}^*$-algebra representation of transformations
\footnote{Work presented at the conference {\em Quantum Theory: Reconsideration of Foundations, 4} 
held on 11-16 June 2007 at the International Centre for Mathematical Modeling in Physics, Engineering and
Cognitive Sciences, V\"axj\"o University, Sweden.}
} 
\author{Giacomo Mauro D'Ariano}{ address={{\em QUIT} Group,
 Dipartimento di Fisica ``A. Volta'', via Bassi 6,
 I-27100 Pavia, Italy, {\em http://www.qubit.it}}}
\begin{abstract} 
  It is shown how a C${}^*$-algebra representation of the transformations of a physical system can
  be derived from two operational postulates: 1) the existence of {\em dynamically independent
    systems}; 2) the existence of {\em symmetric faithful states}. Both notions are crucial for the
  possibility of performing experiments on the system, in preventing remote instantaneous influences
  and in allowing calibration of apparatuses. The case of Quantum Mechanics is thoroughly analyzed.
  The possibility that other no-signaling theories admit a C${}^*$-algebra formulation is discussed.
\end{abstract}
\maketitle
\section{Introduction}
 
In a set of recent papers \cite{darianoVax2006,beyondthequantum,QCM07} I showed how it is possible
to derive the mathematical formulation of Quantum Mechanics in terms of complex Hilbert spaces and
C${}^*$-algebras, starting from a small set of purely operational Postulates concerning experimental
accessibility. In the present manuscript I will focus on C${}^*$-algebra, showing how a
C${}^*$-algebra representation of the transformations of a physical system can be derived from two
operational postulates only, concerning the existence of: 1) dynamically independent systems; 2)
symmetric faithful joint states of two identical systems. Both postulates are crucial for the
possibility of performing experiments, the former preventing uncontrollable remote instantaneous
influences, the latter allowing calibration of experimental apparatuses.

The C${}^*$-algebra representation of the transformations is derived from the postulates via a
Gelfand-Naimark-Segal (GNS) construction \cite{GNS} based on the Jordan decomposition of the
symmetric faithful state. The whole construction holds for finite dimensions, but it is valid also
for infinite dimensions with the proviso that the Jordan decomposition exists on the Banach space of
effects. The notion of {\em adjoint} of a transformation stems from that of faithful state, and
generally depends on it, thus leading to different C${}^*$-algebra representations. On the other
hand, the two postulates together imply that the linear space of "effects" of two identical
independent systems is the tensor product structure of the spaces of the component systems.  

A thorough analysis will show that for the case of Quantum Mechanics the adjoint is actually
independent on the faithful state. However, as it will discussed in the conclusions, the
C${}^*$-algebra representation of transformations is not sufficient to derive Quantum+Classical
Mechanics, as for the customary operator algebras over Hilbert spaces, and in order to select this
case additional postulates are needed. Possible candidates for such postulates, along with the
possibility for other no-signaling theories to admit a C${}^*$-algebra representation of
transformations, are discussed at the end of the paper.

\section{The postulates} 
The general premise of the present axiomatization is the fact that one performs experiments to
gather information on the {\em state} of an {\em object} physical system, and the knowledge of such state
will then enable to predict the results of forthcoming experiments. Moreover, since we necessarily
work with only partial {\em a priori} knowledge of both system and experimental apparatus, the rules
for the experiment must be given in a probabilistic setting. Then, an {\em experiment} on an object
system consists in making it interact with an apparatus, producing one of {\em a set of possible
 transformations} of the object, each one occurring with some probability. Information on the state
of the object at the beginning of the experiment is gained from the knowledge of which
transformation occurred, which is the "outcome" of the experiment signaled by the apparatus. For
the above reasons we can logically {\em identify the experiment with a set of probabilistic
transformations}.
\par We can now introduce the two postulates.
\begin{postulate}[Independent systems]\label{p:independent} There exist independent physical systems.
\end{postulate}
\begin{postulate}[Symmetric faithful state]\label{p:faith} For every composite system made of two identical
 physical systems there exist a symmetric joint state that is both dynamically and preparationally
 faithful.
\end{postulate}
\medskip

\section{The statistical and dynamical structure\label{s:stat}}
The starting point of the axiomatization is the identification {\bf experiment }$\equiv${\em set of
 transformations} that can occur on the object. The apparatus signals which transformation $\tA_j$
of the set $\AA:=\{\tA_j\}$ actually occurs. Now, since the knowledge of the state of a physical
system allows us to predict the results of forthcoming experiments on the object, then it will
allow us to evaluate the probability of any possible transformation in any conceivable experiment.
Therefore, by definition, a {\bf state} $\omega$ of a system is a rule providing probabilities of
transformation, and $\omega(\tA)$ is the probability that the transformation $\tA$ occurs. We
clearly have the completeness $\sum_{\tA_j\in\AA}\omega(\tA_j)=1$, and assume $\omega(\tI)=1$ for
the identical transformation $\tI$, corresponding to adopting $\tI$ as the free evolution (this is
the {\em Dirac picture}, \ie a suitable choice of the lab reference frame). In the following for a
given physical system we will denote by $\Stset$ the set of all possible states and by $\Trnset$ the
set of all possible transformations.
\par When composing two transformations $\tA$ and $\tB$, the probability $p(\tB|\tA)$ that $\tB$
occurs conditional on the previous occurrence of $\tA$ is given by the rule for conditional
probabilities $p(\tB|\tA)=\omega(\tB\circ\tA)/\omega(\tA)$. This sets a new probability rule
corresponding to the notion of {\bf conditional state} $\omega_\tA$ which gives the probability that
a transformation $\tB$ occurs knowing that the transformation $\tA$ has occurred on the object in
the state $\omega$, namely $\omega_\tA\doteq\omega(\cdot\circ\tA)/\omega(\tA)$ \footnote{M. Ozawa
 noticed that the definition of conditional state needs to assume that 
$$\sum_{\tB_j\in\AB}\omega(\tB_j\circ\tA)=\omega(\tA),\qquad\forall\AB,\;\forall\tA.$$
Such assumption which seems not implicit in the present axiomatization, would correspond to a kind
of {\em ``no-signaling from the future''}. It is presently under consideration if this must be
considered as an additional postulate. Notice that such assumption seems to be needed whenever a
notion of conditional state is considered which involves transformations of the system. In the
present context the notion of conditional state is intimately related to that of ``effect'' and to
the action of transformations over effects.} (in the following the central dot ``$\cdot$'' will
always denote the pertinent variable). We can see that the notion of ``state'' itself logically
implies the identification {\em evolution}$\equiv${\em state-conditioning}, entailing a {\em linear
 action of transformations over states} (apart from normalization)
$\tA\omega:=\omega(\cdot\circ\tA)$: this is the same concept of {\bf operation} that we have in
Quantum Mechanics, which gives the conditioning $\omega_\tA=\tA\omega/\tA\omega(\tI)$. In other
words, this is the analogous of the Schr\"{o}dinger picture evolution of states in Quantum Mechanics
(clearly such identification of evolution as state-conditioning also includes the deterministic case
$\tU\omega=\omega(\cdot\circ\tU)$ of transformations $\tU$ with
$\omega(\tU)=1\,\forall\omega\in\Stset$---the analogous quantum channels, including unitary
evolutions. 

From the state-conditioning rule it follows that we can define two complementary types of
equivalences for transformations: {\em dynamical} and {\em informational}. The transformations
$\tA_1$ and $\tA_2$ are {\bf dynamically equivalent} when $\omega_{\tA_1}=\omega_{\tA_2}$
$\forall\omega\in\Stset$, whereas they are {\bf informationally equivalent} when
$\omega(\tA_1)=\omega(\tA_2)$ $\forall\omega\in\Stset$. The two transformations are then completely
equivalent (write $\tA_1=\tA_2$) when they are both dynamically and informationally equivalent,
corresponding to the identity $\omega(\tB\circ\tA_1)=\omega(\tB\circ\tA_2)$,
$\forall\omega\in\Stset,\;\forall\tB\in\Trnset$. We call {\bf effect} the informational equivalence
class of transformations\footnote{This is the same notion of ``effect'' introduced by Ludwig
  \cite{Ludwig-axI}}. In the following we will denote effects with the underlined symbols $\cA$,
$\cB$, etc., or as $[\tA]_\eff$, and we will write $\tA_0\in\cA$ meaning that "the transformation
$\tA$ belongs to the equivalence class $\cA$", or "$\tA_0$ has effect $\cA$", or "$\tA_0$ is
informationally equivalent to $\tA$". Since, by definition one has $\omega(\tA)\equiv\omega(\cA)$,
we will legitimately write $\omega(\cA)$ instead of $\omega(\tA)$.  Similarly, one has
$\omega_\tA(\tB)\equiv \omega_\tA(\cB)$, which implies that
$\omega(\tB\circ\tA)=\omega(\cB\circ\tA)$, leading to the chaining rule
$\cB\circ\tA\in\underline{\tB\circ\tA}$ corresponding to the "Heisenberg picture" evolution of
transformations acting on effects (notice how transformations act on effects from the right). Now,
by definitions effects are linear functionals over states with range $[0,1]$, and, by duality, we
have a convex structure over effects, and we will denote their convex set as $\Cntset$. An {\bf
  observable} is just a complete set of effects $\AL=\{l_i\}$ of an experiment $\AA=\{\tA_j\}$,
namely one has $l_i=\underline{\tA_j}$ $\forall j$ (clearly, one has the completeness relation
$\sum_il_i=1$\footnote{With a little notational abuse sometimes we identify $\cI\equiv 1$, \ie the
  identity effect with the constant functional equal to 1.}). We will call the observable
$\AL=\{l_i\}$ {\bf informationally complete} when each effect $l$ can be written as a real linear
combination $l=\sum_ic_i(l)l_i$ of elements of $\AL$, and when these are linearly independent we
will call the informationally complete observable {\em minimal}.\footnote{In previous literature the
  existence of informationally complete observable has been taken as a postulate. However, in the
  present context it is easy to show that it is always possible to construct a minimal
  informationally complete observable starting from a set of available experiments. The proof is by
  induction, and runs as follows. \def\Ex{{\mathbb E}} By definition there must exists a spanning
  set for $\Cntset_\Reals=\Span_\Reals(\Cntset)$ that is contained in the convex hull $\Cntset$ of
  available effects. The maximal number of elements of this set that are linearly independent will
  constitute a {\em basis}, which we suppose has finite cardinality equal to $\dim(\Cntset_\Reals)$.
  It remains to be shown that it is possible to have a basis with sum of elements equal to 1, and
  that such basis is obtained operationally starting from the available observables from which we
  constructed $\Cntset$.
\par If all observables are {\em uninformative} (\ie with all constant effects $\propto\cI$) , then
$\Cntset_\Reals=\Span_\Reals(\cI)$, $\cI$ is a minimal infocomplete observable, and the statement of the
theorem is proved. Otherwise, there exists at least an observable $\Ex=\{l_i\}$ with $n\geq 2$
linearly independent effects. If this is the only observable, again the theorem is proved.
Otherwise, take a new binary observable $\Ex_2=\{x,y\}$ from the set of available ones (you can take
different binary observables out of a given observable with more than two outcomes by summing up
effects to yes-no observables). If $x\in\Span_\Reals(\Ex)$ discard it. If $x\not\in\Span_\Reals(\Ex)$, then
necessarily also $y\not\in\Span_\Reals(\Ex)$ [since if there exists coefficients $\lambda_i$ such that
$y=\sum_i\lambda_i l_i$, then $x=\sum_i(1-\lambda_i) l_i$]. Now, consider the observable
$$\Ex'=\left\{\tfrac{1}{2}y,\tfrac{1}{2}(l_1+x),\tfrac{1}{2}l_2,\ldots,l_n\right\}$$
(which operationally corresponds to the random choice between the observables $\Ex$ and $\Ex_2$ with
probability $\frac{1}{2}$, and with the events corresponding to $x$ and $l_1$ made
indistinguishable). This new observable has now $|\Ex'|=n+1$ linearly independent effects (since $y$ is
linearly independent on the $l_i$ and one has $y=\sum_{i=1}^n l_i-x=\sum_{i=2}^n l_i +l_1-x$). By
iterating the above procedure we reach $|\Ex'|=\dim(\Cntset_\Reals)$, and we have so realized an
apparatus that measures a minimal informationally complete observable.\qed}
\par The fact that we necessarily work in the presence of partial knowledge about both object and
apparatus corresponds to the possibility of incomplete specification of both states and
transformations, entailing: a) the convex structure on states; b) the addition rule for {\bf
  coexistent transformations}, \ie for transformations $\tA_1$ and $\tA_2$ for which
$\omega(\tA_1)+\omega(\tA_2)\leq 1,\;\forall\omega\in\Stset$ (\ie transformations that can in
principle occur in the same experiment). The addition of the two coexistent transformations is the
transformation $\tS=\tA_1+\tA_2$ corresponding to the event $e=\{1,2\}$ in which the apparatus
signals that either $\tA_1$ or $\tA_2$ occurred, but does not specify which one. Such transformation
is uniquely determined by the informational and dynamical classes as $\forall\omega\in\Stset$:
$\omega(\tA_1+\tA_2)=\omega(\tA_1)+\omega(\tA_2),\; (\tA_1+\tA_2)\omega=\tA_1\omega+ \tA_2\omega$.
The composition "$\circ$" of transformations is distributive with respect to the addition "$+$". We
can also define the multiplication $\lambda\tA$ of a transformation $\tA$ by a scalar
$0\leq\lambda\leq 1$ as the transformation dynamically equivalent to $\tA$, but occurring with
rescaled probability $\omega(\lambda\tA)=\lambda\omega(\tA)$. Now, since for every couple of
transformations $\tA$ and $\tB$ the transformations $\lambda\tA$ and $(1-\lambda)\tB$ are coexistent
for $0\leq\lambda\leq 1$, the set of transformations also becomes a convex set. Moreover, the
transformations make a {\em monoid} (\ie a semigroup with identity), since the composition
$\tA\circ\tB$ of two transformations $\tA$ and $\tB$ is itself a transformation, and there exists the
identical transformation $\tI$ satisfying $\tI\circ\tA=\tA\circ\tI=\tA$ for every transformation
$\tA$.  Therefore, the set of physical transformations $\Trnset$ is a convex monoid.

It is obvious that we can extend the notions of coexistence, sum and multiplication by a scalar from
transformations to effects via equivalence classes. In this way also effects make a convex set. As
an additional step we can extend the convex monoid of physical transformations $\Trnset$ to a real
algebra $\Trnset_\Reals$ by taking differences of physical transformations, and multiply them by
scalars $\lambda>1$. We will call the elements of $\Trnset_\Reals/\Trnset$ {\bf generalized
  transformations}. Likewise, we can introduce {\bf generalized effects}, and denote their linear
space as $\Cntset_\Reals$. On generalized effects we can introduce the norm
$\n{\cA}:=\sup_{\omega\in\Stset}|\omega(\cA)|$, which allows us to introduce also a norm for
transformations as $\n{\tA}:=\sup_{\Cntset_\Reals\ni\n{\cB}\leq 1}\n{\cB\circ\tA}=
\sup_{\Cntset_\Reals\ni\n{\cB}\leq 1}\;\sup_{\omega\in\Stset}\; \omega(\cB\circ\tA)$. Closure in the
respective norm topologies make $\Cntset_\Reals$ a real Banach space and $\Trnset_\Reals$ a real
Banach algebra.\footnote{An algebra of maps over a Banach space can always be made itself a Banach
  space, also satisfying the bound $\n{\tB\circ\tA}\leq\n{\tB}\n{\tA}$ defining a Banach
  algebra. This is true for both the real and the complex cases.\label{foot}}
\par A purely dynamical notion of {\bf independent systems} coincides with the possibility of
performing local experiments. More precisely, we say that two physical systems are {\em independent}
if on the two systems 1 and 2 we can perform {\bf local experiments} $\AA^{(1)}$ and $\AA^{(2)}$,
\ie whose transformations commute each other (\ie
$\tA^{(1)}\circ\tB^{(2)}=\tB^{(2)}\circ\tA^{(1)},\;\forall \tA^{(1)}\in\AA^{(1)},\,\forall
\tB^{(2)}\in\AB^{(2)}$). Notice that the above definition of independent systems is purely
dynamical, in the sense that it does not contain any statistical requirement, such as the existence
of factorized states. The present notion of dynamical independence is so minimal that it can be
satisfied not only by the quantum tensor product, but also by the quantum direct sum
\cite{meta-quantum_trieste06}. Nevertheless, in Sect. \ref{s:dynind} a dimensionality analysis will
show that, in conjunction with the existence of faithful states, dynamical independence agrees only
with the quantum tensor product \footnote{As shown in Refs.
  \cite{darianoVax2006,meta-quantum_trieste06} the tensor product can be derived from the additional
  Postulate stating the {\em local observability principle.}}. In Ref. \cite{meta-quantum_trieste06}
it is shown how the sole dynamical independence implies the impossibility of istantaneous signaling:
the no-signaling condition is crucial for experimental control.

In the following, when dealing with more than one independent system, we will denote local
transformations as ordered strings of transformations as 
$\tA,\tB,\tC,\ldots:=\tA^{(1)}\circ\tB^{(2)}\circ\tC^{(3)}\circ\ldots$. For effects one has the
locality rule $([\tA]_\eff,[\tB_\eff)\in[(\tA,\tB)]_\eff$. The notion of independent systems now
entails the notion of {\em local state}---the equivalent of partial trace in Quantum Mechanics. For
two independent systems in a joint state $\Omega$, we define the {\bf local state} $\Omega|_1$ (and
similarly $\Omega|_2$) as the probability rule $\Omega|_1(\tA)\doteq\Omega(\tA,\tI)$ of the joint
state $\Omega$ with a local transformation $\tA$ acting only on system $1$ and with all other
systems untouched.

\section{The C${}^*$-algebra of transformations} 
Now that we have a real algebra of generalized transformations and a linear space of generalized
effects we want to introduce a positive bilinear form over them, by which we will be able to
introduce a scalar product via the GNS construction \cite{GNS}. The role of such bilinear form will
be played by a {\em faithful state}.

\par We say that a state $\Phi$ of a bipartite system is {\bf dynamically faithful} for system 1
when for every transformation $\tA$ the map $\tA\leftrightarrow(\tA,\tI)\Phi$ is one-to-one, namely
$\forall\tA\in\Trnset_\Reals$ $(\tA,\tI)\Phi=0\Longleftrightarrow\tA=0$. This means that for every
bipartite effect $\cB$ one has $\Phi(\cB\circ(\tA,\tI))=0\Longleftrightarrow\tA=0$. On the other
hand, we will call a state $\Phi$ of a bipartite system {\bf preparationally faithful} for system 1
if every joint bipartite state $\Psi$ can be achieved by a suitable local transformation $\tT_\Psi$
on system 1 occurring with nonzero probability, \ie $\Psi=(\tT_\Psi,\tI)\Phi$, with
$\tT_\Psi\in\Trnset^+$, $\Trnset^+$ denoting the positive cone generated by transformations.
Clearly a bipartite state $\Phi$ that is preparationally faithful is also {\em locally}
preparationally faithful, namely every local state $\psi$ of system 2 can be achieved by a suitable
local transformation $\tT_\psi$ on system 1.
\par In Postulate \ref{p:faith} we also use the notion of {\bf symmetric joint state}. This is
simply defined as a joint state of two identical systems such that for any couple of effects $\cA$
and $\cB$ one has $\Phi(\cA,\cB)=\Phi(\cB,\cA)$. Clearly, for a symmetrical state the notions of
dynamical and preparational faithfulness hold for both systems 1 and 2. 

For a {\em faithful} bipartite state $\Phi$, the {\bf transposed transformation} $\transpPhi{\tA}$
of the transformation $\tA$ is the generalized transformation which when applied to the second
component system gives the same conditioned state and with the same probability as the
transformation $\tA$ operating on the first system, namely $(\tA,\tI)\Phi=(\tI,\transpPhi{\tA})\Phi$
or, equivalently $\Phi(\cB\circ\tA,\cC)=\Phi(\cB,\cC\circ\transpPhi{\tA})$ $\forall
\cB,\cC\in\Cntset$. Clearly the transposed is unique, due to injectivity of the map
$\tA\leftrightarrow(\tA,\tI)\Phi$, and it is easy to check the axioms of transposition
($\transpPhi{\tA+\tB}= \transpPhi{\tA}+\transpPhi{\tB}$, $\transpPhi{\transpPhi{\tA}}=\tA$,
$\transpPhi{\tA\circ\tB}= \transpPhi{\tB}\circ\transpPhi{\tA}$) and that $\transpPhi{\tI}=\tI$.

The main ingredient of a GNS construction for representing transformations would be a positive form
$\varphi$ over transformations based on a notion of adjoint $\tA\to\tA^\dag$ by which one can
construct a scalar product as $\<\tA|\tB\>:=\varphi(\tA^\dag\circ\tB)$ in terms of which we 
have $\<\tA|\tC\circ\tB\>=\<\tC^\dag\circ\tA|\tB\>\equiv\varphi(\tA^\dag\circ\tC\circ\tB)
=\varphi((\tC^\dag\circ\tA)^\dag\circ\tB)$.\footnote{It is not easy to devise a positive form over
  generalized transformations $\Trnset_\Reals$ such that the transposition plays the role of the
  adjoint on a real Hilbert space. Indeed, if we take $\varphi$ as the local state of a symmetric
  faithful state $\varphi=\Phi|_2\equiv\Phi|_1$ we have
  $\varphi(\transpPhi{\tA}\circ\tB)=\Phi(\transpPhi{\tA},\transpPhi{\tB})$, but the fact that $\Phi$
  is positive over the convex set $\Trnset$ of physical transformations doesn't guarantee that its
  extension to generalized transformations $\Trnset_\Reals$ is still positive.} We can extract from
$\Phi$ a positive bilinear form over $\Cntset_\Reals$ (notice that the bilinear form $\Phi$ is
actually defined on effects) using its {\bf Jordan decomposition} in terms of its absolute value
$|\Phi|:=\Phi_+-\Phi_-$. Indeed, the absolute value can be defined thanks to the fact that $\Phi$ is
real symmetric, whence it can be diagonalized over $\Cntset_\Reals$ in the finite dimensional case.
Upon denoting by $\map{P}_\pm$ the orthogonal projectors over the linear space corresponding to
positive and negative eigenvalues, respectively, \footnote{The existence of the orthogonal space
  decomposition corresponding to positive and negative eigenvalues is guaranteed for finite
  dimensions. For infinite dimensions $\Phi$ is just a symmetric form over a real Banach space---the
  space $\Cntset_\Reals$ of generalized effects---and the existence of such decomposition needs to
  be proven.} one has $|\Phi|(\cA,\cB)=\Phi(\varsigma_\Phi(\cA),\cB)$, where
$\varsigma_\Phi(\cA):=(\map{P}_+-\map{P}_-)(\cA)$. The map $\varsigma_\Phi$ is an involution, namely
$\varsigma_\Phi^2=\tI $. The fact that the state is also preparationally faithful implies that the
bilinear form is {\em strictly} positive \cite{darianoVax2006} (namely $|\Phi|(\cA,\cA)=0$ implies
that $\cA=0$). The involution $\varsigma_\Phi$ over $\Cntset_\Reals$ corresponds to a generalized
transformation $\tZ _\Phi\in\Trnset_\Reals$ defined as $\cA\circ\tZ _\Phi:=\varsigma_\Phi(\cA)$,
whence it can be extended to generalized transformations $\Trnset_\Reals$ via
$\cB\circ\varsigma_\Phi(\tA)=\varsigma_\Phi(\varsigma_\Phi(\cB)\circ\tA)$, corresponding to
$\varsigma_\Phi(\tA)=\tZ _\Phi \circ\tA\circ\tZ_\Phi$. Since $\tZ_\Phi^2=\tI$ the extension of
$\varsigma_\Phi$ to $\Trnset_\Reals$ is composition-preserving, \ie
$\varsigma_\Phi(\tB\circ\tA)=\varsigma_\Phi(\tB)\circ\varsigma_\Phi(\tA)$.

The explicit form of $\tZ_\Phi$ can be obtained in terms of the basis $\{f_j\}$ for $\Cntset_\Reals$
reducing the bilinear symmetric form $\Phi$ over $\Cntset_\Reals$ to the canonical form
\begin{equation}
\Phi(f_i,f_j)=s_i\delta_{ij},
\end{equation}
where $s_i=\pm 1$ is the signature of the eigenvector $f_i$. Then one has 
\begin{equation}
\varsigma_\Phi(\cA)=\cA\circ\tZ_\Phi=\sum_j\Phi(f_j,\cA)f_j.
\end{equation}
One can see that $\tau_\Phi\varsigma_\Phi=\varsigma_\Phi\tau_\Phi$. In fact, due to the
symmetry of $\Phi$, $\transpPhi{\tZ_\Phi}=\tZ_\Phi$, since for any couple of elements $f_k,f_l$ of the
basis 
\begin{equation}
\Phi(f_k\circ\transpPhi{\tZ_\Phi},f_l)=\Phi(f_k,f_l\circ\tZ_\Phi)=\Phi(f_l\circ\tZ_\Phi,f_k)=
\delta_{lk}=\Phi(f_k\circ\tZ_\Phi,f_l).
\end{equation}
whence
\begin{equation}
\begin{split}
\transpPhi{\varsigma_\Phi(\tA)}=&
\transpPhi{\tZ_\Phi\circ\tA\circ\tZ_\Phi}=
\transpPhi{\tZ_\Phi}\circ\transpPhi{\tA}\circ\transpPhi{\tZ_\Phi}\\=&
\tZ_\Phi\circ\transpPhi{\tA}\circ\tZ_\Phi=
\varsigma_\Phi(\transpPhi{\tA}).
\end{split}
\end{equation}
We now define the {\bf adjoint map} $\adPhi:=\varsigma_\Phi\tau_\Phi=\tau_\Phi\varsigma_\Phi$. Here
in the following we will also temporarily use the more compact notation $\tA^\dag:=\adPhi(\tA)$,
keeping in mind that the definition of the adjoint generally depends on the faithful state $\Phi$
with respect to which it is defined. Since $\varsigma_\Phi$ is composition preserving whereas
$\tau_\Phi$ is a transposition, one has $(\tB\circ\tA)^\dag=\tA^\dag\circ\tB^\dag$. Moreover, for
$\varphi=\Phi|_1$ we have that $\varphi(\tA^\dag\circ\tB)=|\Phi|(\tau_\Phi(\tA),\tau_\Phi(\tB))$ is
a positive bilinear form over transformations (strictly positive over effects, \ie
$|\Phi|(\cA,\cA)=0\,\Rightarrow\cA=0$ ), and can be used to define a scalar product over
transformations as follows
\begin{equation}\label{eq:scal}
{}_\Phi\!\<\tA|\tB\>_\Phi:=\varphi(\tA^\dag\circ\tB)= \Phi(\varsigma_\Phi\transpPhi{\tA},
\transpPhi{\tB}).
\end{equation}
We can then verify that $\tA^\dag:=\varsigma_\Phi\transpPhi{\tA}$ works as an adjoint for such
scalar product, namely one has
${}_\Phi\!\<\tC^\dag\circ\tA|\tB\>_\Phi={}_\Phi\!\<\tA|\tC\circ\tB\>_\Phi$. In this way
$\varsigma_\Phi$ is identified as the {\bf complex conjugation}, and as usual the adjoint
$\tA^\dag:=\varsigma_\Phi\transpPhi{\tA}=\tau_\Phi\varsigma_\Phi(\tA)$ is the composition of the
transposition with the complex conjugation. Now, by taking complex linear combinations of
generalized transformations and defining $\varsigma_\Phi(c\tA)=c^*\varsigma_\Phi(\tA)$ for
$c\in\Cmplx$, we can extend the adjoint to complex linear combinations of generalized
transformations, whose linear space will be denoted by $\aA\equiv \Trnset_\Cmplx$, which is a
complex algebra. On the other hand, we can trivially extend the real linear space of generalized
effects $\Cntset_\Reals$ to a complex linear space $\Cntset_\Cmplx$ by taking complex linear
combinations of generalized effects. The remaining setting up of the C${}^*$-algebra representation
of $\aA$ is just standard GNS construction, starting from the scalar product between transformations
in Eq. (\ref{eq:scal}). Symmetry and positivity imply the bounding \cite{darianoVax2006}
${}_\Phi\!\<\tA|\tB\>_\Phi\leq\n{\tA}_\Phi\n{\tB}_\Phi$, where we introduced the norm induced by the
scalar product $\n{\tA}_\Phi^2\doteq{}_\Phi\!\<\tA|\tA\>_\Phi$. By taking the equivalence classes
$\aA/\aI$ with respect to the zero-norm elements $\aI\subseteq\aA$ we thus obtain a complex
pre-Hilbert space equipped with a symmetric scalar product, and, since the scalar product is
strictly positive over generalized effects, the elements of $\aA/\aI$ are indeed the generalized
effects, \ie $\aA/\aI\simeq\Cntset_\Cmplx$ as linear spaces. Being endowed with the scalar product
(\ref{eq:scal}) $\aA/\aI$ becomes a pre-Hilbert space, whose completion
$\sH_\Phi:=\overline{\aA/\aI}$ under the norm induced by the scalar product is then a Hilbert space.
In the following we will conveniently denote the equivalence class of transformations containing
$\tA$ in $\aA/\aI$ by the Dirac vector itself $|\tA\>_\Phi\in\sH_\Phi$. From the bounding for the
scalar product it follows that the set $\aI\subseteq\aA$ of zero norm elements $\tX\in\aA$ is a left
ideal (\ie $\tX\in\aI$, $\tA\in\aA$ implies $\tA\circ\tX\in\aI$), whence using our scalar product
defined as in Eq. (\ref{eq:scal}) we can represent elements of $\aA$ (\ie generalized complex
transformations, since $\aA\equiv\Trnset_\Cmplx$) as operators over the pre-Hilbert space of effects
$\Cntset_\Cmplx$. The product in $\aA$ defines the action of $\aA$ on the vectors in $\aA/\aI$, by
associating to each element $\tA\in\aA$ the linear operator $\pi_\Phi(\tA)$ defined on the dense
domain $\aA/\aI\subseteq\sH_\Phi$ as $\pi_\Phi(\tA)|\tB\>_\Phi\doteq|\tA\circ\tB\>_\Phi$. The fact
that $\aA$ is a Banach algebra\footnote{Indeed norms introduced in Sect.  \ref{s:stat} can be
  extended to the respective complex linear spaces, and the norm completion makes $\aA$ also a
  complex Banach algebra, as explained in the footnote \ref{foot}.} also implies that the domain of
definition of $\pi_\Phi(\tA)$ can be easily extended to the whole $\sH_\Phi$ by continuity. Being
now an operator algebra over a complex Hilbert space, $\aA$ becomes a C${}^*$-algebra. We just need
to introduce the norm on transformations as the respective operator norm over $\sH_\Phi$, namely
$\n{\tA}_\Phi:=\sup_{\upsilon\in\sH_\Phi,\n{\upsilon}_\Phi\leq 1}\n{\tA\upsilon}_\Phi$, and
completion of $\aA$ under the norm topology will give a C$^*$-algebra (\ie a complex Banach algebra
satisfying the identity $\n{\tA^\dag\circ\tA}=\n{\tA}^2$), as it can be easily proved by standard
techniques \cite{darianoVax2006}.

I want to emphasize that even though $\sH_\Phi\simeq\Cntset_\Cmplx$ as linear spaces, the elements
$|\tA\>_\Phi\in\sH_\Phi$ should be regarded as element of the dual space of $\Cntset_\Cmplx$, in the
sense that the action of transformations over vectors $|\tA\>_\Phi\in\sH_\Phi$ is from the left---as in
the Schr\H{o}dinger picture---instead of being from the right---as in the Heisenberg picture, \eg
$\pi_\Phi(\tC)|\tA\>_\Phi=|\tC\circ\tA\>_\Phi=$, or
${}_\Phi\<\tA|\pi_\Phi(\tC)={}_\Phi\<\tC^\dag\circ\tA|$, as it follows from the identity
$\<\tB|\pi_\Phi(\tC)|\tA\>_\Phi=\<\tB|\tC\circ\tA\>_\Phi= \<\tC^\dag\circ\tB|\cA\>_\Phi$. The
Schr\H{o}dinger picture is obtained thanks to the transposition in the definition of the scalar
product ${}_\Phi\!\<\tB|\tA\>_\Phi=|\Phi|(\transpPhi{\tB},\transpPhi{\tA})$.

From the definition of the scalar product, and using the fact that the state $\Phi$ is also
preparationally faithful according to Postulate \ref{p:faith}, the Born rule can be written in the
GNS representation as $\omega(\cA)={}_\Phi\<\tA^\dag|\varrho\>_\Phi$, with representation of state
$\varrho=\transpPhi{\cT_\omega}/\Phi(\cT_\omega,\tI)$ \cite{darianoVax2006}, $\tT_\omega$ denoting
the transformation on system 2 corresponding to the local state $\omega$ on system 1, namely
$\omega\propto\Phi(\cdot,\tT_\omega)$. Then, the representation of transformations is
\begin{equation}
\omega(\cB\circ\tA)={}_\Phi\<\tB^\dag|\tA|\rho\>_\Phi:=
{}_\Phi\<\tB^\dag|\tA\circ\rho\>_\Phi.
\end{equation}

\subsection{Connecting two faithful states\label{s:2fait}}
Suppose that $\Omega$ is a symmetric state which is faithful both preparationally and dynamically,
and that $\Phi$ is another such kind of state. Then, there must exists an invertible generalized
transformation $\tF$ in the positive cone $\Trnset^+$ generated by physical transformations, such
that
\begin{equation}
\Phi=(\tF,\tI )\Omega.
\end{equation}
In fact, since $\Omega$ is preparationally faithful, there must exists a local physical
transformation which transforms $\Omega$ into any state with some probability. On the other hand,
since $\Omega$ is dynamically faithful, in order to have also $\Phi$ so, the correspondence between
any other joint state and a local map applied to $\Phi$ must be one-to-one, which is true iff $\tF$
is invertible. If the map $\tF^{-1}$ is itself in the positive cone $\Trnset^+$ generated by
physical transformations, then the state is also preparationally faithful, and viceversa. Indeed,
any pure joint state $\Sigma$ must be written as $\Sigma=(\tS,\tI )\Omega$ with 
$\tS\in\Trnset^+$. Therefore $\Sigma$ can also be obtained probabilistically from $\Phi$ as $(\tilde{\tS
},\tI )\Phi$ using a transformation $\tilde{\tS }\propto\tF\tS \tF^{-1}\in\Trnset^+$ belonging to
the convex cone $\Trnset^+$ generated by physical transformations. Finally, as
regards symmetry, the state $\Phi$ is symmetric iff $\tau_\Omega(\tF)=\tF$, since
\begin{equation}
\begin{split}
\Phi(\tA,\tB)=&\Omega(\tA\circ\tF,\tB)=\Omega(\tB,\tA\circ\tF)=\Omega(\tB\circ\tau_\Omega(\tF),\tA),
\\\Phi(\tA,\tB)=&\Phi(\tB,\tA)=\Omega(\tB\circ\tF,\tA),\quad\forall\tB,\tA\in\Trnset
\end{split}
\end{equation}
and using preparational faithfulness of $\Omega$ one can see that the above identity holds true iff
$\tau_\Omega(\tF)=\tF$ (we remind that two transformations $\tA_1$ and $\tA_2$ are equal iff
$\omega(\tB\circ\tA_1)=\omega(\tB\circ\tA_2)$ $\forall \omega\in\Stset$ and $\forall\tB\in\Trnset$).
Notice now that $\transpOm{\tF^{-1}}=\tF^{-1}$, since $\tI =\transpOm{\tF^{-1}}\circ
\transpOm{\tF}=\transpOm{\tF^{-1}}\circ\tF$.

The transposed with respect to $\Phi$ is obtained as follows
\begin{equation}
(\tA,\tI )(\tF,\tI )\Omega=(\tI,\transpPhi{\tA })(\tF,\tI )\Omega=
(\tI,\transpPhi{\tA }\circ\transpOm{\tF})\Omega
\end{equation}
namely $\transpPhi{\tA }\circ\transpOm{\tF}=\transpOm{\tF}\circ\transpOm{\tA }$, which means that
\begin{equation}\label{tfhi}
\transpPhi{\tA }=\transpOm{\tF^{-1}\circ\tA\circ\tF}=\tF\circ\transpOm{\tA }\circ\tF^{-1}
\end{equation}
The canonical basis of eigenvectors $\{f_j\}$ of the bilinear form $\Phi$ must satisfy the identities
\begin{equation}\label{2form}
s_j\delta_{ij}=\Phi(f_i,f_j),\quad
\delta_{ij}=\Phi(\varsigma_\Phi(f_i),f_j)=|\Phi|(f_i,f_j),
\end{equation}
and upon multiplying by $f_j$ and summing over the index $j$ one obtains
$f_j\circ\tZ_\Phi=\sum_j\Phi(f_i,f_j)f_j$, and since $\{f_i\}$ is a basis for $\Cntset_\Reals$, one as
the identity
\begin{equation}
\cA\circ\tZ_\Phi=\sum_j\Phi(\cA,f_j)f_j,\quad\forall \cA\in\Cntset_\Reals.
\end{equation}
For any couple of elements of the complete basis $\{f_j\}$ for $\Cntset_\Reals$ one has
\begin{equation}
\delta_{ij}=|\Phi|(f_i,f_j)=\Phi(f_i\circ\tZ_\Phi,f_j)=
\Omega(f_i\circ\tZ_\Phi\circ\tF,f_j),
\end{equation}
and since $\{f_i\}$ is a basis for $\Cntset_\Reals$, this corresponds to the identity
\begin{equation}\label{Z1}
\tZ_\Phi\circ\tF\circ\tZ_{\Omega,{\bf f}}=\tI,
\end{equation}
where
\begin{equation}
\cA\circ\tZ_{\Omega,{\bf f}}:=\sum_j\Omega(\cA,f_j)f_j,\quad\forall \cA\in\Cntset_\Reals.
\end{equation}
The definition of $\tZ_{\Omega,{\bf f}}$ generalizes that of $\tZ_\Phi$ in specifying the basis
${\bf f}:=\{f_j\}$ which is generally non canonical for $\Omega$. For ${\bf o}:=\{o_j\}$ canonical for
$\Omega$ one has simply $\tZ_{\Omega}\equiv\tZ_{\Omega,{\bf o}}$.
Upon multiplying by $f_j$ and summing over the index $j$ in Eq. (\ref{2form}) we obtain
\begin{equation}
s_if_i=\sum_j\Phi(f_i,f_j)f_j=\sum_j\Omega(f_i\circ\tF,f_j)f_j=f_i\circ\tF\circ\tZ_{\Omega,{\bf f}}.
\end{equation}
This corresponds to
\begin{equation}
\tZ_\Phi=\tF\circ\tZ_{\Omega,{\bf f}},
\end{equation}
which, in conjunction with Eq. (\ref{Z1}), is a restatement of the involutive nature of $\tZ_\Phi$,
\ie $\tZ_\Phi\circ\tZ_\Phi=\tI$, corresponding also to the identities
\begin{equation}
\tZ_{\Omega,{\bf f}} \circ\tF\circ\tZ_{\Omega,{\bf f}}=\tF^{-1},\quad\tF\circ\tZ_{\Omega,{\bf f}} \circ\tF=\tZ_{\Omega,{\bf f}}^{-1}.
\end{equation}
Therefore, one also has
\begin{equation}
\tZ_\Phi=\tF\circ\tZ_{\Omega,{\bf f}}=\tZ_{\Omega,{\bf f}}^{-1}\circ\tF^{-1}.
\end{equation}
The complex conjugation obeys the symmetry $\transpPhi{\tZ_\Phi}=\tZ_\Phi $ which is needed for a
proper definition of the adjoint. Indeed
\begin{equation}
\transpPhi{\tZ_\Phi}=\tF\circ\transpOm{\tZ_\Phi}\circ\tF^{-1}=
\tF\circ\tZ_{\Omega,{\bf f}}\circ\tF\circ\tF^{-1}=\tF\circ\tZ_{\Omega,{\bf f}} =\tZ_\Phi.
\end{equation}
One has $\transpOm{\tZ_{\Omega,{\bf f}}}=\tZ_{\Omega,{\bf f}}$, since
\begin{equation}
\begin{split}
\Omega(f_i\circ\tZ_{\Omega,{\bf f}},f_j)=&\sum_k\Omega(f_i,f_k)\Omega(f_k,f_j)=
\sum_k\Omega(f_j,f_k)\Omega(f_k,f_i)\\=&\Omega(f_j\circ\tZ_{\Omega,{\bf f}},f_i)=\Omega(f_i,f_j\circ\tZ_{\Omega,{\bf f}})
\end{split}
\end{equation}
We now evaluate the adjoint
\begin{equation}
\adPhi(\tA):=\varsigma_\Phi\transpPhi{\tA}=\tZ_\Phi\circ\tF\circ\transpOm{\tA}\circ\tF^{-1}
\circ\tZ_\Phi=\tZ_{\Omega,{\bf f}}^{-1}\circ\transpOm{\tA}\circ\tZ_{\Omega,{\bf f}},
\end{equation}
and one has
\begin{equation}\label{2ad}
\adPhi\equiv\adOm:=(\cdot)^\dag\quad\Leftrightarrow\quad \tZ_{\Omega,{\bf f}}\equiv\tZ_\Omega,
\end{equation}
namely if $\tZ_\Omega\equiv\tF^{-1}\circ\tZ_\Phi$.
In such case we will also have
\begin{equation}
\tF^{-1}=\varsigma_\Omega(\tF)=\tF^\dag.
\end{equation}

\section{Dynamical independence and tensor product \label{s:dynind}} 
As already mentioned, our notion of dynamical independence---\ie the possibility of performing local
experiments---can be satisfied not only by the quantum tensor product, but also by the quantum
direct sum. This is shown in detail in Ref. \cite{meta-quantum_trieste06}. Here I will show how
Postulate 2---the existence of dynamically and preparationally faithful states---in conjunction with
dynamical independence, leads to the right dimension for the convex set of states of two independent
identical systems according to the tensor product rule.

The state-effect duality leads to the identity $\dim(\Cntset)=\dim(\Stset)+1$,\footnote{For convex
  sets ${\mathfrak C}$, one has $\dim({\mathfrak C}):=\dim\Span({\mathfrak C})$, where
  $\dim\equiv\dim_\Reals$ (if not otherwise stated, the convex sets are always considered real).}
(we remind that one dimension is blocked by state normalization). Then, the existence of a
preparationally and dynamically faithful state guarantees that generalized transformations and
generalized joint states are isomorphic as linear spaces, whence $\dim(\Trnset)=\dim(\Stset^{\times
  2})+1$, $\Stset^{\times 2}$ denoting the set of bipartite states of two identical systems, each
with set of states $\Stset$. Finally, the GNS construction represents generalized transformations as
operators over the Hilbert space of generalized effects, whence $\dim(\Trnset)=\dim(\Cntset)^2$,
from which it follows that $\dim(\Stset^{\times 2})+1=(\dim(\Stset)+1)^2$. Therefore one has
$\dim(\Cntset^{\times 2})=(\dim\Cntset)^2$, and $\dim_\Cmplx(\Cntset_\Cmplx^{\times
  2})=(\dim\Cntset_\Cmplx)^2$ (since $\dim_\Cmplx\Cntset_\Cmplx=\frac{1}{2}\dim\Cntset_\Cmplx=
\dim\Cntset_\Reals$), whence $\Cntset^{\times 2}_\Reals\equiv\Cntset^{\otimes 2}_\Reals$ and
$\Cntset^{\times 2}_\Cmplx\equiv\Cntset^{\otimes 2}_\Cmplx$.  The last identities hold in Quantum
Mechanics, as a consequence of the tensor product of complex Hilbert spaces.

\section{The Quantum C${}^*$-algebra of transformations}
In the following, for given fixed orthonormal basis $\{|j\rangle\}$ for $\sH$ we will denote by $A^*
=\sum_{ij}A_{ij}^*|i\rangle\langle j|$ the operator corresponding to the complex conjugated matrix
of $A=\sum_{ij}A_{ij}|i\rangle\langle j|$, and consistently $\transp{A}=(A^*)^\dag$ will denote the
transposed-matrix operator. With the double ket we denote bipartite vectors $|\Psi\kk\in{\cal
  H}\otimes{\cal H}$, which, keeping the basis $\{|j\rangle\}$ as fixed, are in one-to-one
correspondence with matrices as $|\Psi\kk=\sum_{ij}\Psi_{ij}|i\rangle\otimes|j\rangle$. We will
denote the generalized transformation and the corresponding quantum linear map by the same letter,
and we will do so also for state and its corresponding quantum density operator. Moreover, we will
write composition of quantum maps as $\tB\tA$ as usual, instead of using the operational notation
$\tB\circ\tA$.  In Quantum Mechanics physical transformations correspond to quantum operations (\ie
trace non-increasing completely positive (CP) maps), effects correspond to positive contractions,
generalized transformations $\Trnset_\Reals$ to differences of CP maps, and generalized effects
$\Cntset_\Reals$ to selfadjoint operators. In the following we will denote by $P_\tA $ the positive
operator describing the effect of the quantum operation $\tA $. For example, we will write
\begin{equation}
\rho(\tA)=\Tr[\tA (\rho)]=\Tr\left[\sum_nA_n\rho A_n^\dag\right]=\Tr[\rho P_\tA],\qquad
P_\tA:=\sum_nA_n^\dag A_n.
\end{equation}
We will also use the notation $\tA ^\dag=\sum_nA_n^\dag\cdot A_n$ for the usual adjoint map of $\tA
=\sum_nA_n\cdot A_n^\dag$, and $\transp{\tA }=\sum_n \transp{A_n}\cdot A_n^*$ for the transposed
map. 

I will now construct explicitly the C$^*$-algebra $\Trnset_\Cmplx$ of $c$-generalized
transformations for a general faithful symmetric quantum state $\Phi$. I first consider the case of
the canonical maximally entangled state $\Omega$, and then analyze the general case of faithful
symmetric state.
\subsection{The maximally entangled state of a qudit}

The canonical maximally entangled state of a qudit
\begin{equation}\label{Omega}
\Omega=d^{-1} |I\kk\bb I|,
\end{equation}
is faithful, both dynamically and preparationally. The fact that it is dynamically faithful is just
the Choi-Jamiolowski representation of CP maps. On the other hand, any pure joint state
$d^{-\frac{1}{2}}|S\kk$ can be written as $(S\otimes I)d^{-\frac{1}{2}}|I\kk$ with $d^{-1}\Tr[S^\dag
S]=1$, $\tS\propto S\cdot S^\dag$ quantum operation (\ie $\tS\in\Trnset^+$), whence $\Omega$ is
preparationally faithful. The state $\Omega$ is also symmetric, since for any couple of generalized
effects one has
\begin{equation}
\Omega(\tA,\tB)=\Tr[\tA \otimes\tB (\Omega)]=
\tfrac{1}{d}\Tr[P_\tA \transp{P}_\tB]=\tfrac{1}{d}\Tr[P_\tB \transp{P}_\tA]=\Omega(\tB,\tA).
\end{equation}
The transposition $\tau_\Omega$ is just the customary transposition
$\tau_\Omega\equiv\transp{(\cdot)}$ with respect to any fixed basis $\{|i\>\}$ such that $\Omega$
has all probability amplitudes equal to $d^{-\frac{1}{2}}$. Indeed, it is easy to check that
\begin{equation}
(\tA \otimes\tI )(|I\kk\bb I|)=(\tI \otimes\transp{\tA })(|I\kk\bb I|).
\end{equation}

In order to construct an eigenbasis for the Jordan form, consider the following selfadjoint
operators
\begin{equation}
X_{kl}=\tfrac{1}{\sqrt{2}}(|k\>\< l|+|l\>\< k|),\quad Y_{kl}=\tfrac{i}{\sqrt{2}} (|k\>\< l|-|l\>\< k|),\quad
k<l,\qquad Z_l=|l\>\<l|.
\end{equation}
One has
\begin{equation}
\Tr[X_{kl}X_{k'l'}]=\delta_{kl'}\delta_{lk'}+\delta_{kk'}\delta_{ll'}\equiv \delta_{kk'}\delta_{ll'},
\end{equation}
since for $k=l'>k'$ one has $k'=l>k$. Similarly we have 
$\Tr[Y_{kl}Y_{k'l'}]=\delta_{kk'}\delta_{ll'}$, and $\Tr[Z_kZ_{k'}]=\delta_{kk'}$, and, moreover 
\begin{equation}
\Tr[X_{kl}Y_{k'l'}]=\Tr[Z_lY_{k'l'}]=\Tr[Z_lX_{k'l'}]=0.
\end{equation}
Therefore, the following is a canonical basis for the Jordan form of $\Phi$
\begin{equation}
\begin{split}
[C_j]=&\left[Z_0,Z_1,\ldots, Z_{d-1}, X_{01},X_{02},\ldots,X_{0,d-1},\right.\\ & \left. X_{12},X_{13},\ldots,
X_{1,d-1}, \ldots X_{d-2,d-1},Y_{01},\ldots,Y_{d,d-1}\right],
\end{split}
\end{equation}
with Jordan form
\begin{equation}
\Omega(C_i,C_j)=\Tr[C_iC_j^*]=\delta_{ij}s_j,
\end{equation}
$Y_{kl}$ ($0\leq k<l\leq d-1$) spanning the eigenspace with negative eigenvalue of the symmetric form
$\Omega$. It follows that the transformation $\varsigma_\Omega$ corresponds to the complex conjugation
$\varsigma_\Omega\equiv (\cdot)^*$ with respect to the same fixed orthonormal basis $\{|i\>\}$ used
for transposition. We can construct the Kraus form for the corresponding generalized transformation
$\tZ_\Omega$, passing through the construction of the corresponding Choi-Jamiolowski operator
\begin{equation}\label{choiE}
\tZ_\Omega\otimes\tI(|I\kk\bb I|)=\tZ_\Omega\otimes\tI\left(
\sum_j C_j^*\otimes C_j\right)=\sum_jC_j\otimes C_j=\sum_j C_j^\dag\otimes C_j=E,
\end{equation}
which is just the unitary swap operator $E$, with eigenvectors 
\begin{equation}\label{eigenE}
E|C_j\kk=|C_j^*\kk=s_j|C_j\kk,
\end{equation}
corresponding to the Kraus form for the generalized transformation $\tZ$ 
\begin{equation}
\tZ=\sum_js_j C_j\cdot C_j.
\end{equation}
\par The GNS representation of transformations over effects is provided by the following scalar product
\begin{equation}
{}_\Omega\<\tA|\tB\>_\Omega:=\tfrac{1}{d}\<\!\!\<I|\check{A}^\dag\check{B}|I\>\!\!\>,
\end{equation}
where corresponding to the map $\tA =\sum_nA_n\cdot A_n^\dag$ we define the operator
$\check{A}:=\sum_n A_n\otimes A_n^*$ such that $\check{A}|X\kk=|\tA (X)\kk$. 
Indeed, we can check the identities 
\begin{equation}\label{scalOm}
{}_\Omega\<\tA|\tB\>_\Omega:=\Omega(\tA^\dag,\tB')=
\tfrac{1}{d}\Tr\left[\sum_mA_mA^\dag_m\sum_nB_nB^\dag_n\right]=\tfrac{1}{d}
\Tr[P_{\tA^\dag}P_{\tB^\dag}],
\end{equation}
\begin{equation}
{}_\Omega\<\tA|\tC\circ\tB\>_\Omega:=\Phi(\tA^\dag,\tB'\circ\tC')=
\tfrac{1}{d}\bb P_{\tA^\dag}|\check{C}|P_{\tB^\dag}\kk
=\tfrac{1}{d}\Tr[P_{\tA^\dag}\tC (P_{\tB^\dag})]
\end{equation}
Explicitely, the GNS representation of transformation over effects is 
\begin{equation}
|\tA\>=\check{A}|I\>\!\!\>=|\tA(I)\>\!\!\>=|P_{\tA^\dag}\kk,\quad \tB|\tA\>=\check{B}|\tA (I)\kk
=|\tB\tA (I)\kk=|\tB (P_{\tA ^\dag})\kk.
\end{equation}
For qubits the canonical Jordan basis, will be given by the set of four Pauli matrices
$\sigma_0\equiv I,\sigma_x,\sigma_y,\sigma_z$ normalized as $C_j=\tfrac{1}{\sqrt2}\sigma_j$,
corresponding to the Jordan form
\begin{equation}
\Omega(C_i,C_j)\doteq \tfrac{1}{2}\Tr[\sigma_i\sigma_j^*]=
\begin{bmatrix}
1 & 0& 0& 0\\
0 & 1& 0& 0\\
0 & 0& -1& 0\\
0 & 0& 0& 1\\
\end{bmatrix}:=\delta_{ij}s_j.
\end{equation}
Here $\sigma_y$ spans the eigenspace with negative eigenvalue of $\Omega$.

\subsection{General faithful state}
According to subsection \ref{s:2fait} the general form of a joint faithful state of two identical quantum
systems with finite dimensional Hilbert space $\sH$ can be always recast in the following way
\begin{equation}
\Phi=\tfrac{1}{d}\sum_l|F_l\kk\bb F_l|=(\tF\otimes\tI )\Omega
\end{equation}
with $\Omega$ given in Eq. (\ref{Omega}), and $\tF=\sum_lF_l\cdot F_l^\dag$ invertible CP map (not
necessarily trace-non-increasing), and with $\tF^{-1}$ also CP (normalization corresponds to
$\Tr[\sum_lF_l^\dag F_l]=d$, $d=\dim(\sH)$).  Moreover, the state $\Phi$ is symmetric iff
$\transp{(\tF^{-1})}=\tF^{-1}$, corresponding to the operator identity $E\Phi E=\Phi$, $E$ denoting
the swap operator. According to
Eq. (\ref{tfhi}) the transposed with respect to $\Phi$ is given by
\begin{equation}
\transpPhi{\tA }=\tF\transp{\tA }\tF^{-1}
\end{equation}
The canonical basis of eigenvectors $\{C_j\}$ of the bilinear form $\Phi$ must satisfy the identity
\begin{equation}\label{2form}
s_j\delta_{ij}=\Phi(C_i,C_j)=\Tr[(C_i\otimes C_j)\sum_l|F_l\kk\bb F_l|]=
\Tr[\transp{C_j}\tF^\dag(C_i)].
\end{equation}
Upon multiplying by $C_j$ and summing over the index $j$ in Eq. (\ref{2form}) we obtain
\footnote{In the present quantum context the notation $\tF^\dag(X)$ corresponds to the Heisenberg
 picture $\cX\circ\tF$, with $X$ selfadjoint operator representing the generalized effect $\tX$.}
\begin{equation}
s_iC_i=\sum_j\Tr[\transp{C_j}\tF^\dag(C_i)]C_j=:\tC\tF^\dag(C_i)
\end{equation}
where $\tC:=\sum_j\Tr[\transp{C_j}\cdot]C_j$.
Identity (\ref{2form}) is equally satisfied by the set $\{C_i^\dag\}$ with the same eigenvalue.
Therefore, it is always possible to choose the operators $C_j$ as selfadjoint, and $\tC ^\dag
\equiv\tC $. It is also easy to check that $\transp{\tC}=\tC$, since
\begin{equation}\label{CCE}
\begin{split}
\tC\otimes\tI (|I\kk\bb I|)=&\sum_jC_j\otimes\Tr_1[\transp{C_j}\otimes I|I\kk\bb I|]=
\sum_jC_j\otimes C_j\\=&\sum_j\Tr_2[ I\otimes\transp{C_j}|I\kk\bb I|]\otimes C_j=\tI \otimes\tC(|I\kk\bb I|).
\end{split}
\end{equation}
Using completeness of $\{C_j\}$ and their self-adjointness, it is easy to see that 
\begin{equation}
\tC(X)=\sum_j\Tr[\transp{C_j}X]C_j=\sum_j\Tr[C_j\transp{X}]C_j=
\sum_j\Tr[C_j^\dag\transp{X}]C_j=\transp{X},
\end{equation}
namely
\begin{equation}
\tC=\transp{(\cdot)},
\end{equation}
and using Eq. (\ref{CCE}) one can see that $\sum_j C_j\otimes C_j=E$ and $\{C_j\}$ are
Hilbert-Schmidt orthonormal. Clearly $\tC^2=\map{I}$, $\tC\tM\tC= \tM^*$,
\ie $\tC=\tZ_\Omega$. According to (\ref{2ad}) this will then guarantee that the adjoint will be
independent on the faithful state $\Phi$. The map $\tC\tF^\dag$ acting on $C_i$ gives their
complex conjugated, and since $\{C_i\}$ is a selfadjont basis of the real linear space of
selfadjoint operators, $\tC\tF^\dag$ is the complex conjugation over all selfadjoint
operators, namely \footnote{On the other hand, for a generic self-adjoint operator it is easy to
 check that
$$
\tZ_\Phi (A)=\sum_k\Phi(C_k,A)C_k=\sum_{kl}\Tr[F_l^*A\transp{F_l}\transp{C_k}]C_k
=\sum_{kl}\Tr[F_l^\dag A F_l\transp{C_k}]C_k=\tC\tF^\dag(A).$$}
\begin{equation}
\tZ _\Phi=\tC\tF^\dag=\tF\tC.
\end{equation}
The complex conjugation obeys the symmetry $\transpPhi{\tZ_\Phi}=\tZ_\Phi $ which is needed for a
proper definition of the adjoint. Indeed
\begin{equation}
\transpPhi{\tZ_\Phi}=\tF\transp{\tZ_\Phi}\tF^{-1}=
\tF\tC\tF\tF^{-1}=
\tF\tC=\tZ_\Phi.
\end{equation}
Since, by definition, the map $\tZ $ is involutive, one has
\begin{equation}
\tI =\tF\tC\tF\tC=\tF\tF^*=\tF\tF^\dag
\end{equation}
whence
\begin{equation}
\tF^{-1}=\tF^*=\tF^\dag.
\end{equation}
Finally, the adjoint of a map $\tA $ is just the usual adjoint, since
\begin{equation}
\varsigma_\Phi\transpPhi{\tA }=\tZ _\Phi\tF\transp{\tA }\tF^{-1}\tZ_\Phi=
\tC \transp{\tA }\tC =\transp{\tA }{}^*=\tA ^\dag,
\end{equation}
or, equivalently,
\begin{equation}
\tau_\Phi\varsigma_\Phi(\tA)=\tF\transp{(\tZ_\Phi \tA \tZ_\Phi )}\tF^{-1}=
\tF\tC\tF\transp{\tA }\tC\tF\tF^{-1}=
\tC\transp{\tA }\tC=\tA ^\dag.
\end{equation}
In Table \ref{t:sum} I summarize the most relevant identities and definitions. 
\begin{center}
\begin{table}
\begin{tabular}{|c||c|c|}
\hline
object & definition & identities\\
\hline
\hline
$\Phi$ &$\sum_l|F_l\kk\bb F_l|$ & $E\Phi E=\Phi$ \\ 
\hline
$\tF$ &$\sum_lF_l\cdot F_l^\dag$ & $\tF=\transp{\tF}$, $\tF^{-1}=\tF^*=\tF^\dag$
\\ \hline 
$\tC$ & $\sum_j\Tr[\transp{C_j}\cdot]C_j$ & $\tC=\transp{(\cdot)}=
\tC^\dag=\transp{\tC}$, $\tC\tM \tC=\tM ^*$\\ \hline
$\tZ_\Phi $ & $\sum_j\Phi(C_j,\cdot)C_j$ & $\tZ_\Phi =\tC\tF^\dag=\tF\tC$\\ \hline
$\transpPhi{\cdot}$ && $\transpPhi{\tM }=\tF\transp{\tM }\tF^{-1}$\\ \hline 
$\varsigma_\Phi(\cdot)$&& $\varsigma_\Phi(\tM )=\tZ_\Phi\tA\tZ_\Phi $, $\transpPhi{\tZ_\Phi}=\tZ_\Phi$\\ \hline 
$\adPhi(\cdot)$ && $\transpPhi{\varsigma_\Phi(\tA)}=\varsigma_\Phi(\transpPhi{\tA })=\tA ^\dag$\\ \hline 
${}_\Phi\<\tA|\tB\>_\Phi$ & $\Phi(\tA^\dag,\transpPhi{\tB})$ &
${}_\Phi\<\tA|\tB\>_\Phi=\tfrac{1}{d}\sum_l\bb F_l|\check{A}^\dag\check{B}|F_l\kk$ \\
\hline 
\end{tabular}
\caption{Summary of most relevant identities and definitions\label{t:sum}}
\end{table}
\end{center}
Explicitely, the GNS representation is given by
\begin{equation}\label{explicitGNS}
{}_\Phi\<\tA|\tB\>_\Phi:=\Phi(\tA^\dag,\transpPhi{\tB})=\tfrac{1}{d}\sum_l\bb
F_l|\check{A}^\dag\check{B}|F_l\kk, 
\end{equation}
where for any CP map $\tA=\sum_iA_i\cdot A_i^\dag$ one has $\check{A}:=\sum_iA_i\otimes A_i^*$ (we
remind the normalization $\Tr[\sum_lF_l^\dag F_l]=d$ of state $\Phi$ in terms of the Kraus operators
of $\map{F}$). For trace-preserving $\tF$ one would obtain the same scalar product as in Eq.
(\ref{scalOm}), \ie ${}_\Phi\<\tA|\tB\>_\Phi:=\Tr[P_{\tA^\dag}P_{\tB^\dag}]$, however, since
$\tF^{-1}=\tF^*$ is also trace preserving, the only possibility would be $\tF=U\cdot U^\dag$
unitary, and with the additional constraint $U=\transp{U}$ coming from symmetry of $\Phi$.

\subsection{The most general quantum scalar product}
We start now from the most general scalar product between two quantum transformations and
show that it must be of the form (\ref{explicitGNS}). The most general form of scalar product
between two operators $A$ and $B$ in $\Bnd{H}$ is 
\begin{equation}
\varphi(A^\dag B)=\sum_j\<\upsilon_j|A^\dag B|\upsilon_j\>,\quad 
\sum_j\<\upsilon_j|\upsilon_j\>=1 
\end{equation}
where normalization corresponds to $\varphi(I)=1$. For quantum transformations the most general
scalar product can be constructed upon regarding transformations as operators on $\Bnd{H}$ (in
infinite dimensions, more precisely, as operators on the Hilbert space of the Hilbert-Schmidt
operators). Therefore, upon considering a complete set of operators $\{E_i\}$, one has
\begin{equation}
(\map{B},\map{A})=\sum_i\bb \map{B}(E_i)|\map{A}(E_i)\kk =
\sum_i\bb E_i| \breve{B}^\dag\breve{A}|E_i\kk, \quad \Tr\left(\sum_iE_i^\dag E_i\right)=1,
\end{equation}
which is exactly of the general form (\ref{explicitGNS}). Notice that the general form
(\ref{explicitGNS}) corresponds to a state $\Phi$ that is mixed, being the convex combination 
$\Phi=\sum_i\Tr[F_i^\dag F_i]|\tilde{F}_i\kk\bb\tilde{F}_i|$, with 
$\tilde{F}_i:=F_i/\sqrt{\Tr[F_i^\dag F_i]}$. 

\section{Conclusions}
In conclusion I want to emphasize that the fact that Postulates 1 and 2 imply a $C^*$-algebra
representation for transformations, and with the correct Born-rule pairing and the correct
dimensionality for the tensor-product structure of bipartite systems, is not sufficient to assert
that the only possible theory derived from the postulates is Quantum Mechanics. Indeed, as for the
general $C^*$-algebras of operators on Hilbert spaces, Classical Mechanics is also included as
special case, corresponding to Abelian $\aA$, and, more generally, a combination of both Quantum and
Classical in a direct sum of irreducible algebra representations, such as in the presence of
constant of motions and/or super-selection rules. Indeed preliminary analysis \cite{Shortme} show
that more general theories can satisfy both postulates, such as the non-local no-signaling
probabilistic theories generally referred to as {\em PR boxes} \cite{PR}. This is the case, for
example, of the model in Ref.  \cite{short-2005}, which possesses a symmetric faithful state,
however with $\dim(\Cntset_\Reals)=3$, which cannot be quantum.

As regards additional postulates selecting Quantum Mechanics from the set of theories admitting
C${}^*$-algebras representations, one may adopt Postulate 4 in Ref. \cite{darianoVax2006} concerning
the possibility of achieving an informationally complete observable by means of a perfectly
discriminating observable over system+ancilla. However, such postulate may look quite {\em ad hoc},
being essentially a restatement of existence of Bell measurements (Bell measurements are locally
informationally complete for one system for almost every state-preparation of the other system).
Alternative candidates for the quantum-extracting postulate are under study, considering what is
specific of the quantum C${}^*$-algebra, \eg the fact that in the quantum case the $C^*$-algebra of
transformations $\aA$ is a kind of {\em multiplier algebra} \cite{Murphy} of the $C^*$-algebra
$\Bnd{H}$.

I want to stress that the dimensionality identity in Sect. \ref{s:dynind} concerning only identical
independent systems could be generalized to the case of different systems. This, however, will need
to consider transformations between different systems. Thus, also the symmetry of the faithful state
must be relaxed, upon considering a suitable transformation that maps the largest to the smallest
system.  Finally, the faithfulness condition itself may be relaxed, obtaining a generally unfaithful
C${}^*$-algebra representation. Thus the C${}^*$-algebra representation of transformations will be
just equivalent to the probabilistic framework endowed with the postulated existence of dynamically
independent systems. A complete analysis of this direction will be the subject of a forthcoming
publication \cite{forthcoming}.

\section*{Acknowledgments}
This research has been supported by the Italian Minister of University and Research (MIUR) under
program Prin 2005. I acknowledge very useful discussions with Tomy Short, Dirk Schlingeman, and
Masanao Ozawa. Part of the present work has been done in Cambridge UK, during the workshop
{\em Operational probabilistic theories as foils to quantum theory}, July 2-13, 2007.

\end{document}